\documentclass[prd,preprint,showpacs]{revtex4-1}
\usepackage[utf8]{inputenc}
\usepackage{amsmath}
\usepackage{latexsym}
\usepackage{amsfonts}
\usepackage{graphicx}
\usepackage{mathrsfs}
\usepackage{CJK}
\usepackage{longtable}
\usepackage{booktabs}

\usepackage{hyperref}
\hypersetup{
  colorlinks = true,
  urlcolor = blue,
  linkcolor = blue,
  citecolor = green,
  filecolor = magenta,
}

\newcommand{\pd}{\partial}

\begin{abstract}
In this paper, the polarization contents of Einstein-\ae ther theory and the generalized TeVeS theory are studied.
The Einstein-\ae ther theory has five polarizations, while the generalized TeVeS theory has six.
In particular, transverse and longitudinal breathing polarization are mixed.
The~possibility of using pulsar timing arrays to detect the extra polarizations in Einstein-\ae{}ther theory was also investigated.
The analysis showed that different polarizations cannot be easily distinguished by using pulsar timing arrays in this theory.
For generalized TeVeS theory, one of the propagating modes travels much faster than the speed of light due to the speed bound set by GW170817.
In some parameter subspaces, the strong coupling does not take place, so this theory is excluded.
\end{abstract}

\begin{document}

\title{Gravitational waves in Einstein-\AE ther theory and generalized TeVeS theory after GW170817}
\author{Shaoqi Hou}
\email{shou1397@hust.edu.cn}
\thanks{talk given by this author.}
\affiliation{School of Physics, Huazhong University of Science and Technology, Wuhan, Hubei 430074, China}
\author{Yungui Gong}
\email{yggong@hust.edu.cn}
\thanks{corresponding author.}
\affiliation{School of Physics, Huazhong University of Science and Technology, Wuhan, Hubei 430074, China}

\maketitle

\section{Introduction}

The Laser Interferometer Gravitational-Wave Observatory (LIGO)
Scientific and Virgo collaborations have detected gravitational waves since 14~September 2015 \cite{Abbott:2016blz,Abbott:2016nmj,Abbott:2017vtc,Abbott:2017oio,TheLIGOScientific:2017qsa,Abbott:2017gyy}.
A new era began when it became possible to probe general relativity (GR) through the high speed, strong-field regime.
Among the six detections,
GW170817 was the first binary neutron star merger event, accompanied by gamma-ray burst GRB 170817A  \cite{TheLIGOScientific:2017qsa,Goldstein:2017mmi,Savchenko:2017ffs}.
Observations led to a very stringent constraint on the speed of GWs
, which constrains many alternative theories of gravity.
Alternatives to GR usually predict that GWs have up to four additional polarizations
in addition to the plus and cross ones \cite{Eardley:1974nw,Capozziello:2013wha,Capozziello:2014hia,Will:2014kxa,Liang:2017ahj,Hou:2017bqj,Gong:2017bru}.
Einstein-\ae ther theory \cite{Jacobson:2000xp} and the generalized tensor-vector-scalar (TeVeS) theory \cite{Seifert:2007fr} contain several extra degrees of freedom (d.o.f.), so it predicts many extra polarizations.
The identification of these extra polarizations and their detection are the main topics of this paper.
We also take into account the implications of the existing experimental constraints on these theories, especially including the speed bounds from GW170817.
A gauge-invariant formalism is devised to obtain GW solutions and identify the polarizations.
For a discussion on the GWs of black holes according to Einstein-\ae ther theory, please refer to Refs.~\cite{Konoplya:2006ar,Konoplya:2006rv}.

In this paper, we first present the GW solutions in Einstein-\ae{}ther theory in Section~\ref{sec-einae-gw}, where
we solve the equations of motion.
We thus find the polarization states.
Then, we discuss the constraints on the theory, and after that, the possible detection by pulsar timing arrays.
A similar analysis is also applied to generalized TeVeS theory in Section~\ref{sec-gtvs}.
Throughout this paper, units are chosen such that the speed of light in vacuum is $c=1$.

\section{Gravitational Waves in Einstein-\AE{}ther Theory}\label{sec-einae-gw}

In Einstein-\ae ther theory, gravity is mediated by the metric tensor, $g_{\mu\nu}$, and the normalized timelike \ae{}ther field $u^\mu$.
The action can be found in Ref.~\cite{Jacobson:2004ts}, and there are four parameters, $c_i\,(i=1,2,3,4)$, that measure the coupling between $u^\mu$ and $g_{\mu\nu}$.
Since $u^\mu$ is normalized and timelike, it~defines a~preferred reference frame everywhere in the spacetime, so it breaks the local Lorentz invariance (LLI).
To~obtain GW solutions, the metric and the \ae ther field are perturbed such that $g_{\mu\nu} = \eta_{\mu\nu} +~h_{\mu\nu},\, u^\mu = \underline{u}^\mu + v^\mu$ with $\underline u^\mu = \delta^\mu_0$.
{Then, the linearized equations of motion can be solved to get GW solutions which is the method used to get GW solutions based on the generic theory of gravity.
A gauge is usually fixed so that the equations of motion take the form of waves.
One may also choose to use gauge-invariant variables \cite{Flanagan:2005yc}, as in the present paper.}
The diffeomorphism invariance of the action allows to define the gauge-invariant variables.
First, $h_{\mu\nu}$ and $v^\mu$ are decomposed in the following way:  $v^0=h_{00}/2=\phi,\, v^j=\mu^j+\partial^j\omega,\, h_{tt} = 2\phi, \,h_{tj} = \beta_j+\partial_j\gamma$, and~$h_{jk} = h_{jk}^\mathrm{TT}+H\delta_{jk}/3+\partial_{(j}\epsilon_{k)}+\left(\partial_j\partial_k-\frac{1}{3}\delta_{jk}\nabla^2\right)\rho$ \cite{Flanagan:2005yc}.
Here, $h_{jk}^\mathrm{TT}$ satisfies $\pd^kh_{jk}^\mathrm{TT}=0$ and $\eta^{jk}h_{jk}^\mathrm{TT}=0$.
$\beta_j,\,\epsilon_j$ and $\mu^j$ are transverse vectors.
Nine gauge-invariant variables can be constructed, which are $h_{jk}^\mathrm{TT}$, $\Phi = -\phi+\dot\gamma-\frac{1}{2}\ddot\rho, \,  \Theta = \frac{1}{3}(H-\nabla^2\rho),   \, \Omega=\omega+\frac{1}{2}\dot\rho,\,  \Xi_j = \beta_j-\frac{1}{2}\dot\epsilon_j,\,\text{and }  \Sigma_j=\beta_j+\mu_j$ \cite{Gong:2018cgj}.
Not all of them are propagating.
In fact, solving the equations of motion gives five propagating d.o.f.: $h_{jk}^\text{TT},\, \Sigma_j$ and $\Omega$.
They generally propagate at three different speeds: $s_g,\,s_v$, and $s_s$.

\emph{Polarizations of gravitational waves}:
If the matter fields minimally couple with $g_{\mu\nu}$ only, the~polarization content of GWs is determined by the linearized geodesic deviation equation $\ddot x^j=d^2x^j/dt^2=-R_{tjtk}x^k$ with $x^j$ being the deviation vector.
Assuming the plane GWs propagate in the $+z$ direction, one can easily identify the polarizations. The plus polarization is given by $\hat P_+=-R_{txtx}+R_{tyty}=\ddot h_+$, and the cross polarization is $\hat P_\times=R_{txty}=-\ddot h_\times$.
The vector-$x$ polarization is represented by $\hat P_{xz}=R_{txtz}\propto\partial_3\dot\Sigma_1$,
and the vector-$y$ polarization is $\hat P_{yz}=R_{txty}\propto\partial_3\dot\Sigma_2$.
The transverse breathing polarization is specified by
$\hat P_b=R_{txtx}+R_{tyty}\propto\dddot\Omega$,
and the longitudinal polarization is $\hat P_l=R_{tztz}\propto\dddot\Omega.$
Note that $\Omega$ excites a mixed state of $\hat P_b$ and $\hat P_l$.

\emph{Discussion on the constraints}:
Previous observations have set various constraints on the parameter space of the theory.
These constraints include these on the post-Newtonian parameters, $\alpha_1$ and $\alpha_2$ \cite{Will:2014kxa},  and the requirements that the GW carry positive energy \cite{Jacobson:2008aj}, and there should not be
gravitational Cherenkov radiation~\cite{Elliott:2005va} etc.
Combining all the constraints shows that this theory is highly constrained.
To make more explicit predictions, we picked some specific points in the allowed parameter space, as shown in Table~\ref{tab-spds}.
In the left table, $c_i$s take values such that LLI is respected, while in the right table, LLI is violated.
It is clear that these values are very small which requires severe fine-tuning.

\begin{table}
  \centering
    \caption{The choices of parameters and the corresponding speeds of the vector and scalar GWs. 
    In the left part of the table, $c_i$s are normalized by $10^{-16}$ and $s_g=1+7\times10^{-15}$.
    In the right part of the table, $c_i$s are normalized by $10^{-9}$ and $s_g=1$.}\label{tab-spds}
  {\small\begin{tabular}{cccc|cccc}
    \hline
    $c_1$ & 7.7 & 8.4 & 9.1 & $c_1=-c_3$ & 6.06 & 3.59 & 2.83\\
    $c_2$  &$-$5.5 & $-$6.2 & $-$6.8 & $c_2$ & 3.66 & 2.58 & 2.10\\
    $c_3$ & 6.3 & 5.6 & 4.9  & $c_4$  & $-4.06$ & $-1.59$ & $-0.83$\\
    $c_4$ &$-$5.2  & $-$3.7 & $-$2.6 & & & &\\
   \hline
    $s_v$ & 1.74 & 1.34 & 1.19 &  $s_v$& 1.74&1.34&1.19\\
    $s_s$ & 1.83 & 1.29 & 1.05 &  $s_s$&1.83&1.29&1.05\\
    \hline
  \end{tabular}
}
\end{table}

\begin{figure}
  \centering
  \includegraphics[width=0.4\textwidth]{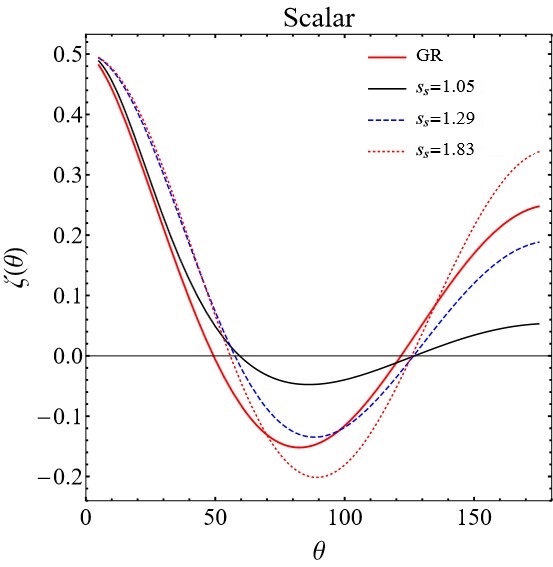}\quad\quad
  \includegraphics[width=0.4\textwidth]{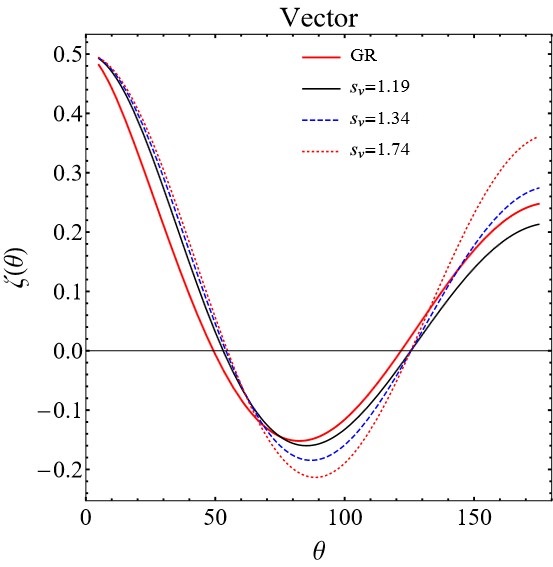}
  \caption{The normalized cross correlation function $\zeta(\theta)$ for the scalar (the left panel) and the vector (the right panel) GWs when $c_i$ have the values shown in the left table in Table~\ref{tab-spds}.}\label{fig-zeta-scl}
\end{figure}

\begin{figure}
  \centering
  \includegraphics[width=0.4\textwidth]{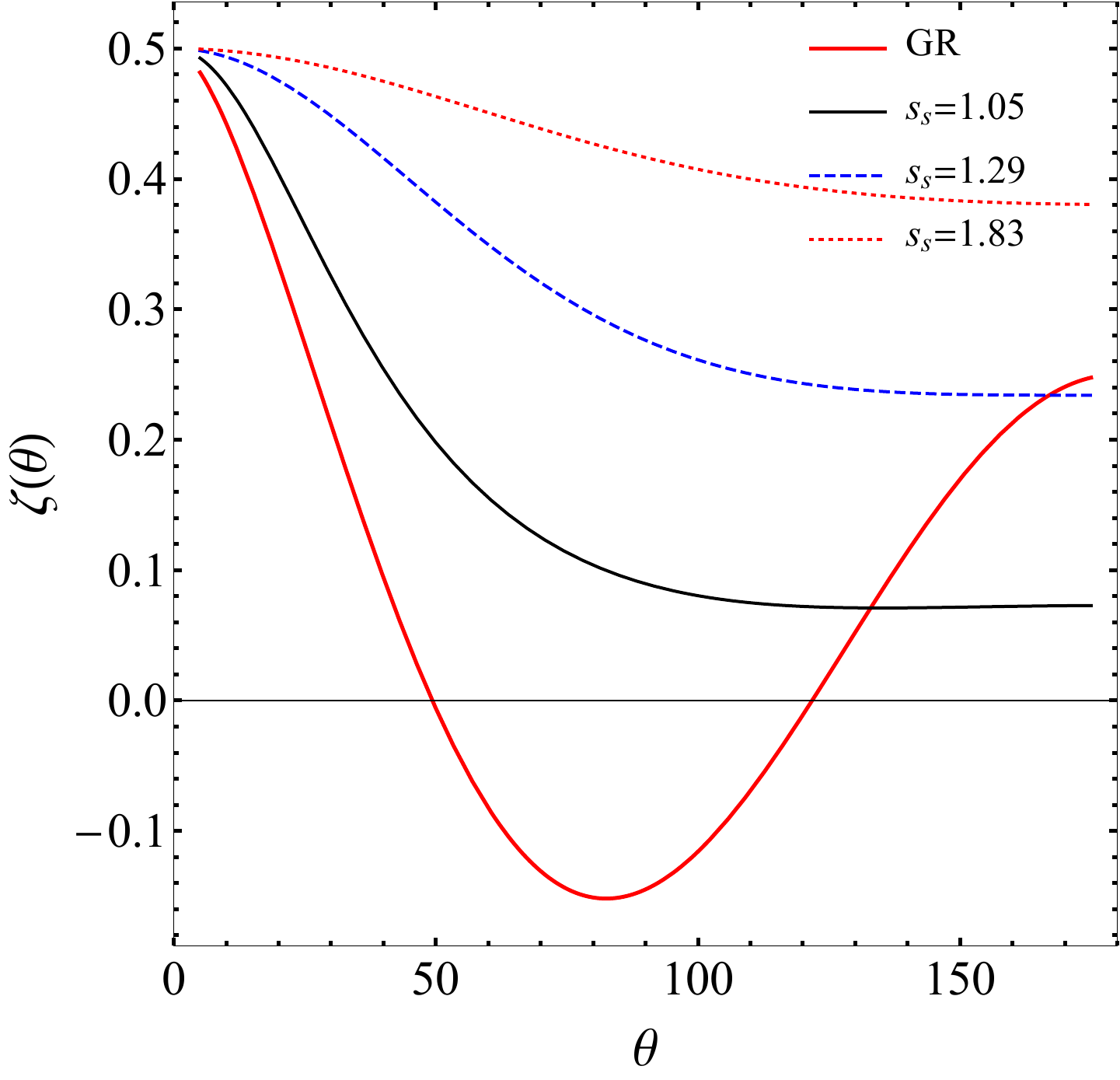}
  \caption{The normalized cross-correlation function $\zeta(\theta)=C_s(\theta)/C_s(0)$ for the scalar GW when the $c_i$s take the values shown in Table~\ref{tab-spds}.}\label{fig-scl-c130}
\end{figure}

\emph{Pulsar timing arrays}:
Pulsar timing arrays (PTAs) can measure the timing residuals $R(t)$ of photons emitted from pulsars.
There is cross correlation between timing residuals for pulsars, which is given by  $C(\theta)=\langle R_a(t)R_b(t)\rangle$,
where $\theta$ is the angular separation between pulsars $a$ and $b$, and the brackets $\langle\,\rangle$ imply the ensemble average over the stochastic GW background.
The functional form of $C(\theta)$ is determined by the type of the polarization of GWs, so PTAs can examine the polarizations of GWs.
For~Einstein-\ae{}ther theory, one can calculate $C(\theta)$, and the results are shown in Figures~\ref{fig-zeta-scl} and \ref{fig-scl-c130}.
Figure~\ref{fig-zeta-scl} shows  the behavior of $\zeta(\theta)=C(\theta)/C(0)$ as a function of $\theta$  for the scalar and the vector polarizations at different speeds corresponding to the values of $c_i$ listed in the left table in Table~\ref{tab-spds}.
The GR's prediction (the red solid curves) is also plotted which approximately represents $\zeta(\theta)$ for the tensor~GW.
It is clear that these curves are very similar to each other, so it would be difficult to distinguish different polarizations.
If one chooses the values for the $c_i$s given in Table~\ref{tab-spds}, the $\zeta(\theta)$ for the scalar GW is modified, as shown in Figure~\ref{fig-scl-c130}.
Since if $c_{13}=0$, there are no vector polarizations, the~corresponding $\zeta(\theta)$ was not plotted.
Figure~\ref{fig-scl-c130} shows that $\zeta(\theta)$ for the scalar GW is very different from the one for the tensor GW.
So, it is easier to distinguish the scalar polarizations from the tensor~ones.

\section{Gravitational Waves in the Generalized TeVeS Theory}\label{sec-gtvs}

\begin{figure}
  \centering
  \includegraphics[width=0.4\textwidth]{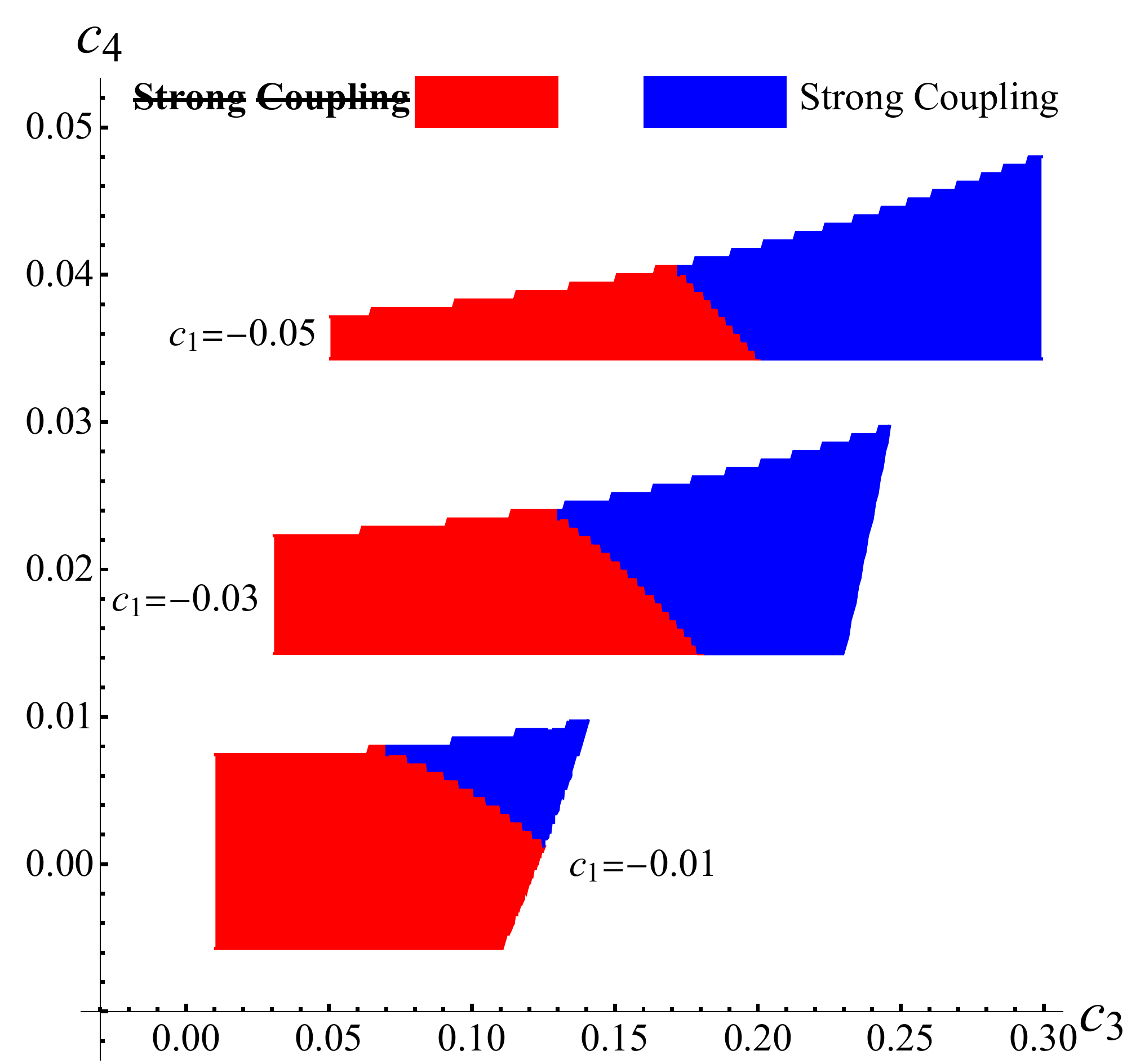}
  \caption{Parameter subspaces (colored areas) allowed by the experimental constraints.
  }\label{fig-sc-nosc}
\end{figure}

Generalized TeVeS theory is the generalization of the theory originally proposed by Bekenstein to attack the dark matter problem \cite{Bekenstein:2004ne}.
Compared with Einstein-\ae{}ther theory, it has an additional scalar field, $\sigma$, which also mediates gravity.
Matter fields minimally couple to the physical metric $\tilde g_{\mu\nu}=e^{-2\sigma}g_{\mu\nu}-2u_\mu u_\nu \sinh(2\sigma)$.
One can easily verify that there are six d.o.f.: $h_{jk}^\text{TT}$, $\Sigma_j$, $\Omega$, and $\sigma$.
They~propagate at four different speeds named $\tilde s_g,\,\tilde s_v,\,\tilde s_s$ and $\tilde s_0$, respectively.
In addition, there are the plus, cross, vector-$x$, and vector-$y$ polarizations.
The two scalar d.o.f.  excite two copies of mixed states of the transverse breathing and the longitudinal polarizations.
There are also several previous experimental constraints on this theory, as given in Refs.~\cite{Bekenstein:2004ne,Sagi:2009kd,Sagi:2007hb,Lasky:2008fs,Lasky:2009sw}.
Combining all of these constraints shows that the speed ($\tilde s_s$) for $\Omega$ is generally much larger than 1, which might lead to the faster decay of binary systems, so this theory might be excluded.
However, a very large speed might result in strong coupling.
We examined whether strong coupling for $\Omega$ takes place.
Figure~\ref{fig-sc-nosc} displays the parameter subspaces that are compatible with the experimental constraints.
Strong coupling does not exist in the red regions,
while in the blue areas, strong coupling takes place.
Then, the generalized TeVeS theory is excluded in the parameter space where strong coupling does not exist.
Further~analysis is required to determine whether it survives in the blue regions.

\section{Conclusions}\label{sec-con}

In this paper, we used the gauge-invariant variable formalism to obtain the linear GW solutions about the Minkowski background and the
polarization contents of Einstein-\ae ther theory and the generalized TeVeS theory.
There are five polarization states in Einstein-\ae ther theory
and six in generalized TeVeS theory.
The longitudinal and transverse breathing modes together form a~single state for the scalar d.o.f. in both theories.
For Einstein-\ae{}ther theory, the analysis showed that it might be difficult for PTAs to distinguish different polarizations when LLI is respected, while it is easier to do so when $s_g=1$ and LLI is violated.
For generalized TeVeS theory, it was found out that it is excluded by the speed bounds on GWs in the parameter regions where the strong coupling of the scalar d.o.f. $\Omega$ does not suffer from strong coupling.

\begin{acknowledgements}
We thank Ted Jacobson for constructive discussions.
This research was supported in part by the Major Program of the National Natural Science Foundation of China under Grant No. 11690021 and the National Natural Science Foundation of China under Grant No. 11475065.
We also thank Cosimo Bambi for the organization of the conference \emph{International Conference on Quantum Gravity} that took place in Shenzhen, China, 26-28 March, 2018. This paper is based on a talk given at the mentioned conference.
\end{acknowledgements}


\end{document}